\documentclass[%
 aip,
 amsmath,amssymb,
 preprint,%
]{revtex4-1}

\usepackage{graphicx}
\usepackage{dcolumn}
\usepackage{bm}

\usepackage[utf8]{inputenc}
\usepackage[T1]{fontenc}
\usepackage{mathptmx}
\usepackage{booktabs}
\usepackage{siunitx}
\usepackage{multirow}
\usepackage{threeparttable}

\newcommand{\eTone}{$e^{T}$ 1.0}
\newcommand{\eT}{$e^{T}$}

\newcommand{\mrm}{\mathrm}

\newcommand{\ket}[1]{\vert #1 \rangle}

\begin{document}

\preprint{AIP/123-QED}

\title[\eTone: an open source electronic structure program]{\eTone: an open source electronic structure program with emphasis on coupled cluster and multilevel methods}

\author{Sarai D.~Folkestad}
\thanks{Equal contributions}
\affiliation{Department of Chemistry, Norwegian University of Science and Technology, 7491 Trondheim, Norway} 
\author{Eirik F.~Kjønstad}
\thanks{Equal contributions}
\affiliation{Department of Chemistry, Norwegian University of Science and Technology, 7491 Trondheim, Norway}
\author{Rolf H.~Myhre}
\thanks{Equal contributions}
\affiliation{Department of Chemistry, Norwegian University of Science and Technology, 7491 Trondheim, Norway}
\author{Josefine H.~Andersen}
\affiliation{DTU Chemistry---Department of Chemistry, Technical University of Denmark, DK-2800 Kongens Lyngby, Denmark}
\author{Alice Balbi}
\affiliation{Scuola normale superiore, Piazza dei Cavalieri, 7, 56126 Pisa PI, Italy}
\author{Sonia Coriani}
\affiliation{DTU Chemistry---Department of Chemistry, Technical University of Denmark, DK-2800 Kongens Lyngby, Denmark}
\author{Tommaso Giovannini}
\affiliation{Department of Chemistry, Norwegian University of Science and Technology, 7491 Trondheim, Norway}
\author{Linda Goletto}
\affiliation{Department of Chemistry, Norwegian University of Science and Technology, 7491 Trondheim, Norway}
\author{Tor S. Haugland}
\affiliation{Department of Chemistry, Norwegian University of Science and Technology, 7491 Trondheim, Norway}
\author{Anders Hutcheson}
\affiliation{Department of Chemistry, Norwegian University of Science and Technology, 7491 Trondheim, Norway}
\author{Ida-Marie Høyvik}
\affiliation{Department of Chemistry, Norwegian University of Science and Technology, 7491 Trondheim, Norway}
\author{Torsha Moitra}
\affiliation{DTU Chemistry---Department of Chemistry, Technical University of Denmark, DK-2800 Kongens Lyngby, Denmark}
\author{Alexander C. Paul}
\affiliation{Department of Chemistry, Norwegian University of Science and Technology, 7491 Trondheim, Norway} 
\author{Marco Scavino}
\affiliation{Scuola normale superiore, Piazza dei Cavalieri, 7, 56126 Pisa PI, Italy}
\author{Andreas S. Skeidsvoll}
\affiliation{Department of Chemistry, Norwegian University of Science and Technology, 7491 Trondheim, Norway} 
\author{Åsmund H. Tveten}
\affiliation{Department of Chemistry, Norwegian University of Science and Technology, 7491 Trondheim, Norway} 
\author{Henrik Koch}
\email{henrik.koch@sns.it}
\affiliation{Department of Chemistry, Norwegian University of Science and Technology, 7491 Trondheim, Norway} 
\affiliation{Scuola normale superiore, Piazza dei Cavalieri, 7, 56126 Pisa PI, Italy}

\date{\today}

\begin{abstract}
    The \eT~program is an open source electronic structure package with emphasis 
    on coupled cluster and multilevel methods. 
    It includes efficient spin adapted implementations of
    ground and excited singlet states, as well as equation of motion oscillator strengths, for CCS, CC2, CCSD, and CC3. 
    Furthermore, 
    \eT~provides unique capabilities such as multilevel Hartree-Fock and multilevel CC2, 
    real-time propagation for CCS and CCSD, 
    and efficient CC3 oscillator strengths. 
    With a coupled cluster code 
    based on an efficient Cholesky decomposition algorithm for the electronic repulsion integrals,
    \eT~has similar advantages as codes using density fitting, 
    but with strict error control.
    Here we present the main features of the program and demonstrate its performance 
    through example calculations.
    Because of its availability, 
    performance, 
    and unique capabilities, 
    we expect \eT~to become a valuable resource to the electronic structure community.
\end{abstract}

\maketitle

\section{Introduction}
During the last five decades, 
a wide variety of models and algorithms have been developed within the field 
of electronic structure theory and many program packages are now available to the community.\citep{helgaker2014molecular}
Programs with extensive coupled cluster functionality include
CFOUR,\citep{stanton2010cfour} 
Dalton,\citep{dalton} 
GAMESS,\citep{GAMESS} 
Molcas,\citep{molcas} 
Molpro,\citep{werner2012molpro}
NWChem,\citep{valiev2010}
ORCA,\citep{orca}
Psi4,\citep{Parrish2017Psi4}
QChem,\citep{shao2015qchem} and
TURBOMOLE.\citep{turbomole}
Although these are all general purpose quantum chemistry programs, 
each code is particularly feature rich or efficient in specific areas. 
For instance,
a large variety of response properties\citep{Helgaker2012response} have been implemented in Dalton,
CFOUR is particularly suited for
gradients\citep{Koch1990gradients,Stanton1993gradients} and geometry optimization,
and QChem is leading in equation of motion\citep{Stanton1993_EOM,Krylov2008EOMreview} (EOM) features.
However, due to the long history of many of these programs,
it can be challenging to modify and optimize existing features or to integrate new methods and algorithms.

In 2016, 
we began developing a coupled cluster code based on Cholesky decomposed electron repulsion integrals.\citep{Beebe1977,Koch2003} 
While starting anew, 
we have drawn inspiration from Dalton\citep{dalton} and used it extensively for testing purposes. 
Our goal is to create an efficient, flexible, and easily extendable foundation upon which coupled cluster methods and features---both established and new---can be developed.
That code has now evolved beyond a coupled cluster code
into an electronic structure program. It is named \eT~after the
expression for the
coupled cluster ground state wave function,\citep{COESTER1960477}
\begin{equation}
    \ket{\Psi} = e^T \ket{\mathrm{R}},
\end{equation}
and released as an open source program licensed under the GNU General Public License 3 (GPL 3.0).

The first version of \eT~offers an optimized Hartree-Fock (HF) code and a wide range of standard coupled cluster methods. 
It includes the most efficient
published implementations of Cholesky decomposition of the electron repulsion integrals\citep{Folkestad2019} and of coupled cluster singles, doubles and perturbative triples\citep{Koch1997,myhre2016multilevel} (CC3).
Furthermore, \eT~features
the first released implementations of 
multilevel HF\citep{saether2017density} (MLHF),
multilevel coupled cluster singles and perturbative doubles\citep{myhre2014mlcc, folkestad2019multilevel} (MLCC2), 
and explicitly time-dependent coupled cluster singles (TD-CCS) and 
singles and doubles (TD-CCSD) theory. 
All coupled cluster models can be used in quantum mechanics/molecular mechanics\cite{warshel1972,levitt1975} (QM/MM) calculations, or be combined with the polarizable continuum model\citep{tomasi2005quantum,mennucci2012polarizable} (PCM).

\eT~is primarily written in modern Fortran, 
using the Fortran 2008 standard. 
The current version of the code is interfaced to two external libraries: 
Libint 2\cite{Valeev2017} for the atomic orbital integrals and 
PCMSolver 1.2\cite{di2019pcmsolver} for PCM embedding.
In addition, 
\eT~applies the 
runtest library\citep{runtest} for testing and 
a CMake module from autocmake\citep{autocmake} 
to locate and configure BLAS and LAPACK.

With the introduction of the 2003 and 2008 standards, Fortran has become an object oriented programming language.
We have exploited this to make \eT~modular, readable, and easy to extend.
Throughout the program,
we use OpenMP\citep{OMP5} 
to parallelize computationally intensive loops.
In order to preserve code quality,
extensive code review and enforcement of a consistent standard has been prioritized from the outset.
While this requires extra effort from both developers and maintainers,
it pays dividends in code readability and flexibility.

\section{Program features}
\subsection{Coupled cluster methods}\label{sec:features:cc_methods}
The \eT~program features all standard coupled cluster methods up to perturbative triples:
singles (CCS),
singles with perturbative doubles\citep{Christiansen1995} (CC2),
singles and doubles\citep{Purvis1982} (CCSD),
singles and doubles with non-iterative perturbative triples\citep{Raghavachari1989} (CCSD(T)),
and singles and doubles with perturbative triples\citep{Koch1997} (CC3).
At the CCSD(T) level of theory, only ground state energies can be computed.
For all other methods,
efficient spin adapted implementations of ground and excited singlet states are available. 
Moreover, 
dipole and quadrupole moments, 
as well as EOM oscillator strengths, 
can be calculated. 
Equation of motion polarizabilities are available at the CCS, CC2, and CCSD levels of theory.

A number of algorithms are implemented to solve the coupled cluster equations.
For linear and eigenvalue equations, 
we have implemented the Davidson method.\cite{DAVIDSON197587} 
This algorithm is used to solve the ground state multiplier equations, 
response equations, 
and excited state equations. 
To handle nonlinear coupled cluster equations, 
we have implemented algorithms that use direct inversion of the iterative subspace\citep{Pulay1982,PULAY1980393} (DIIS) to accelerate convergence. 
The ground state amplitude equations can be solved using DIIS combined with the standard\citep{helgaker2014molecular,SCUSERIA1986236}  quasi-Newton algorithm or exact Newton-Raphson. 
We also use a DIIS-accelerated algorithm\citep{Hattig2000} for the nonlinear excited state equations in CC2 and CC3.
Our implementation of DIIS incorporates the option to use the related conjugate residual with optimal trial
vectors\citep{CROP2008,Ettenhuber2015} (CROP) method for acceleration. 
For the nonperturbative coupled cluster methods, 
the asymmetric Lanczos algorithm is also available.\citep{lanczos1950iteration,Coriani2012}

The time-dependent coupled cluster equations can be explicitly solved for 
CCS and CCSD\citep{Koch1990,Pedersen2019} using Euler,
Runge-Kutta 4 (RK4), 
or Gauss-Legendre (GL2, GL4, GL6) integrators.
This requires implementations of the amplitude and multiplier equations with complex variables. 
Any number of classical electromagnetic pulses can be specified in the length gauge, 
assuming that the dipole approximation is valid.
A modified version of the fast Fourier transform library 
FFTPACK 5.1\citep{SWARZTRAUBER198445} is used to extract frequency domain information.

\subsection{Cholesky decomposition for the electronic repulsion integrals}
Cholesky decomposition is an efficient method
to obtain a compact factorization of the 
rank deficient electron repulsion integral matrix.\citep{Beebe1977,Koch2003,aquilante2011cholesky}
All post HF methods in \eT
rely on the Cholesky vectors to construct the electron repulsion integrals.
One advantage of factorization is the reduced 
storage requirements;
the size of the Cholesky vectors scales as $\mathcal{O}(n_{\mrm{AO}}^3)$
while the full integral matrix scales as $\mathcal{O}(n_{\mrm{AO}}^4)$.
The Cholesky vectors are kept in memory when possible,
but are otherwise stored on disk.
Another advantage is that
they 
allow for efficient construction and transformation of
subsets of the integrals.
The Cholesky decomposition in \eT~is highly efficient,
consisting of a two step procedure that reduces both storage requirements and 
computational cost compared to earlier algorithms.\citep{Folkestad2019} 

\subsection{Hartree-Fock}
The restricted (RHF) and unrestricted HF (UHF) models are implemented in \eT.
The implementations are integral direct and exploit 
Coloumb and exchange screening and permutation symmetry. 
We use a superposition of atomic densities\citep{VanLenthe2006} (SAD) 
initial guess constructed from 
spherically averaged UHF calculations on the constituent atoms. 
The Hartree-Fock equations are solved using a Roothan-Hall self-consistent field (SCF)
algorithm accelerated either by DIIS or CROP.
To improve the screening and reduce the number of integrals that must be evaluated,
density differences are used to construct the Fock matrix.

\subsection{Multilevel and multiscale methods} \label{sec:features:multilevel_methods}
In MLHF, a region of the molecular system is defined as active.
A set of active occupied orbitals are obtained through a restricted, 
partial Cholesky decomposition of an initial idempotent AO
density matrix.\cite{sanchez2010cholesky} 
Active virtual orbitals are obtained by constructing 
projected atomic orbitals\cite{pulay1983localizability,saebo1993local} (PAOs)
centered on the active atoms.
The PAOs are orthonormalized through the canonical orthonormalization procedure.\citep{lowdin1970nonorthogonality}
The MLHF equations are solved using a DIIS accelerated, 
MO based, Roothan-Hall SCF algorithm. 
Only the active MOs are optimized.\citep{hoyvik2019convergence}

The most expensive step of an MLHF calculation 
is the construction of the inactive two electron contribution to the Fock matrix.
As the inactive orbitals are frozen, 
it is only necessary to calculate this term once.
The iterative cost in MLHF is dominated by the construction of the active two electron
contribution to the Fock matrix.
An additional Coulomb and exchange screening, 
which targets accuracy of the matrix in the active MO basis, 
reduces the cost. 
The active orbitals are localized and, consequently,
elements of the AO Fock matrix which correspond 
to AOs distant from the active atoms
will not significantly contribute to the active MO Fock matrix.
This is similar to the screening used in MLHF specific Cholesky decomposition
of the electron repulsion integrals.\citep{Folkestad2019}

In MLCC2,\citep{myhre2013extended,myhre2014mlcc,myhre2016multilevel,folkestad2019multilevel} an active orbital space is treated at the CC2 level of theory, while the remaining inactive orbitals are treated at the CCS level of theory. MLCC2 excitation energies are implemented in \eT.
The active space is constructed using 
either approximated correlated natural transition 
orbitals,\citep{hoyvik2017correlated, baudin2017correlated} 
Cholesky orbitals, 
or Cholesky occupied orbitals and PAOs spanning the virtual space. 

Frozen orbitals are implemented for all coupled cluster methods in \eT. 
In addition to the standard frozen core (FC) approximation,
reduced space coupled cluster calculations can be performed using semi-localized orbitals.
This type of calculation is suited to describe localized properties.
In reduced space calculations, the occupied space is constructed from Cholesky orbitals and PAOs are used to generate the virtual space.

Two QM/MM approaches are available in \eT:
electrostatic QM/MM embedding\cite{senn2009qm} and the polarizable QM/Fluctuating Charge\cite{cappelli2016integrated} (QM/FQ) model. 
In the former, 
the QM density interacts with a set of fixed charges 
placed in the MM part of the system.\cite{senn2009qm}
In QM/FQ,
the QM and MM parts mutually polarize.
Each atom in the MM part has a charge that varies as a response to differences in atomic electronegativities and the QM potential.\cite{cappelli2016integrated} 
These charges enter the QM Hamiltonian through a term that is nonlinear in the QM density.\cite{lipparini2012linear}

PCM embedding can be used in \eT~for an implicit description of the external environment.
A solute is described at the QM level and is placed in a molecule shaped cavity.
The environment is described in terms of an infinite, 
homogeneous,
continuum dielectric
that mutually polarize with the QM part,
as in QM/FQ.\cite{giovannini2019tpa}

In the QM/PCM and QM/FQ implementations, additional terms are only added to the Fock matrix. Additional terms, at the coupled cluster level, can also be considered.\cite{caricato2019ijqc,cammi2009jcp,caricato2012absorption,caricato2013vertical,ren2019coupled}

\subsection{Spectroscopic properties and response methods}
Coupled cluster is one of the most accurate 
methods for modelling spectroscopic properties,
and both UV/Vis and X-ray absorption spectra can be modelled in \eT.
Core excitations are obtained through 
the core valence separation (CVS)
approximation.\citep{Cederbaum1980} 
CVS is implemented as a projection\citep{Coriani2015,Coriani2016erratum} for 
CCS, 
CC2, 
MLCC2, 
and CCSD. 
For CC3, 
amplitudes and excitation vector elements that do not contribute are not calculated. 
This reduces the scaling of the iterative computational cost from 
$\mathcal{O}(n_\mrm{MO}^7)$ to $\mathcal{O}(n_\mrm{MO}^6)$.

Intensities are obtained from EOM oscillator strengths,\citep{Stanton1993_EOM,Krylov2008EOMreview} 
which are available for CCS, 
CC2,
CCSD,
and CC3. 
In addition, 
linear response\citep{Koch1990} (LR) oscillator strengths can be calculated at the CCS level of theory. 
The asymmetric Lanczos algorithm\citep{lanczos1950iteration,Coriani2012} 
can be used to directly obtain both energies and EOM oscillator strengths for CCS, CC2 and CCSD. 
It can also be combined with the CVS approximation.

Real-time propagation offers a nonperturbative approach to model absorption spectra.
Following an initial pulse that excites the system, 
the dipole moment from the subsequent time evolution
can be Fourier transformed to extract the excitation energies and intensities.

Valence ionization potentials are implemented for CCS, 
CC2,
and CCSD. 
A bath orbital that does not interact with the system 
is added to the calculation. 
Excitation vector components not involving this orbital 
are projected out in an approach similar to the projection in CVS.\citep{Coriani2015,Coriani2016erratum}

\section{Illustrative applications and performance tests}
In this section, 
we will demonstrate some of the capabilities of \eT~with example calculations.
Thresholds are defined as follows.
Energy thresholds refer to the change in energy from the previous iteration. The maximum norm of the gradient vector 
is used in Hartree-Fock calculations. 
For coupled cluster calculations in \eT~and Dalton, 
residual thresholds refer to the $L^2$ norm of the residual vectors. 
Finally, the Cholesky decomposition
threshold refers to the largest absolute error on the diagonal of the electron repulsion  integral matrix. This threshold gives an upper bound to the error of all matrix elements.
Coupled cluster calculations were performed with either Cholesky vectors or electron repulsion integrals in memory.
All geometries are available from Ref.~\citenum{eTgeometries}.

\subsection{Coupled cluster methods}

\begin{figure}
    \centering
    \includegraphics[width=0.8\linewidth]{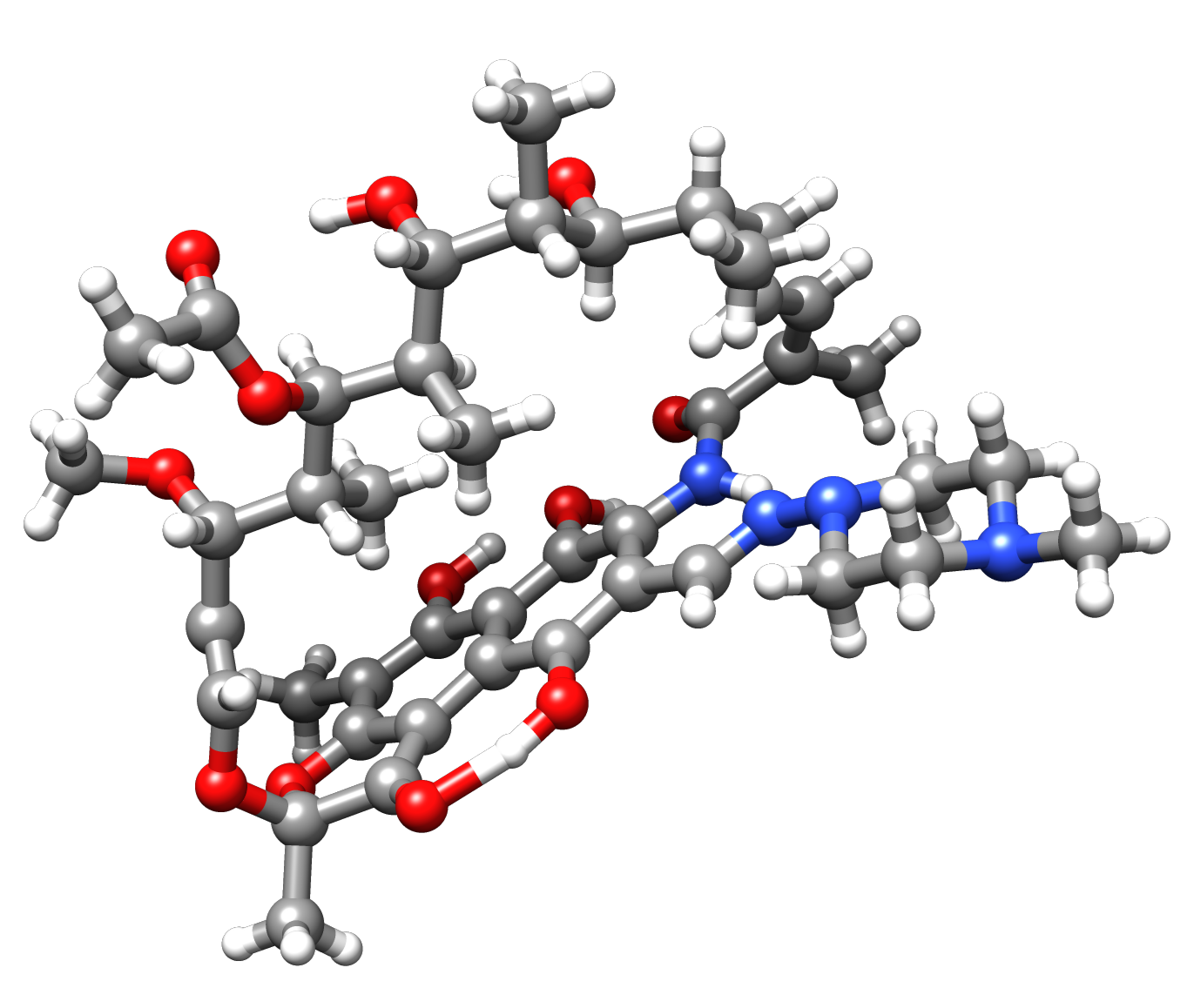}
    \caption{Rifampicin}
    \label{fig:rifampicin}
\end{figure}

\begin{table}
    \centering
    \caption{The lowest FC-CC2/aug-cc-pVDZ excitation energy $(\omega)$ of rifampicin. $n_{\mrm{frozen}}$ is the number of frozen core orbitals.}
    \begin{tabular}{c c c c}
    \toprule
        $n_{\mrm{AO}}$ &
        $n_{\mrm{MO}}$ &
        $n_{\mrm{frozen}}$ &
        $\omega$\\
    \midrule
        1879 & 1865 & 59 &  $\SI{2.5789}{\eV}$ \\
    \bottomrule
    \end{tabular}
    \label{tab:rifampicin}
\end{table}

The CC2 method is known to yield excitation energies with an error of approximately $\SI{0.1}{\eV}$ for states with single excitation character.\citep{Christiansen1996,kannar2014benchmarking,kannar2016accuracy} The iterative cost of CC2 scales as $\mathcal{O}(n_{\mrm{MO}}^5)$ and it may be implemented with an $\mathcal{O}(n_{\mrm{MO}}^2)$ memory requirement.
In Table \ref{tab:rifampicin}, we report the lowest FC-CC2/aug-cc-pVDZ excitation energy of the antibiotic rifampicin\citep{rifampicin} (chemical formula C$_{43}$H$_{58}$N$_{4}$O$_{12}$, see Figure \ref{fig:rifampicin}). 
The calculated excitation energy is $\SI{2.58}{\eV}$, which is 
consistent with the orange color of the compound and
the experimental value of $\SI{2.61}{\eV}$.\citep{BENETTON1998639}
The ground state was converged to a residual threshold of $10^{-6}$, and the excited state was converged to residual and energy thresholds of $10^{-3}$ and $10^{-6}$, respectively. 
We used
a Cholesky decomposition threshold of $10^{-2}$, 
which is sufficient to ensure accuracy of excitation energies in CC2 and CCSD (see Table \ref{tab:lsd_ccsd}).
The calculation was performed on two Intel Xeon Gold 6138 processors, using 40 threads and ${360}$ GB shared memory. 
The average iteration time for the ground state equations was $\SI{73}{\minute}$, 
and the average iteration time for the excited state equations was $\SI{9}{\hour}$.

\begin{figure}
    \centering
    \includegraphics[width=\linewidth]{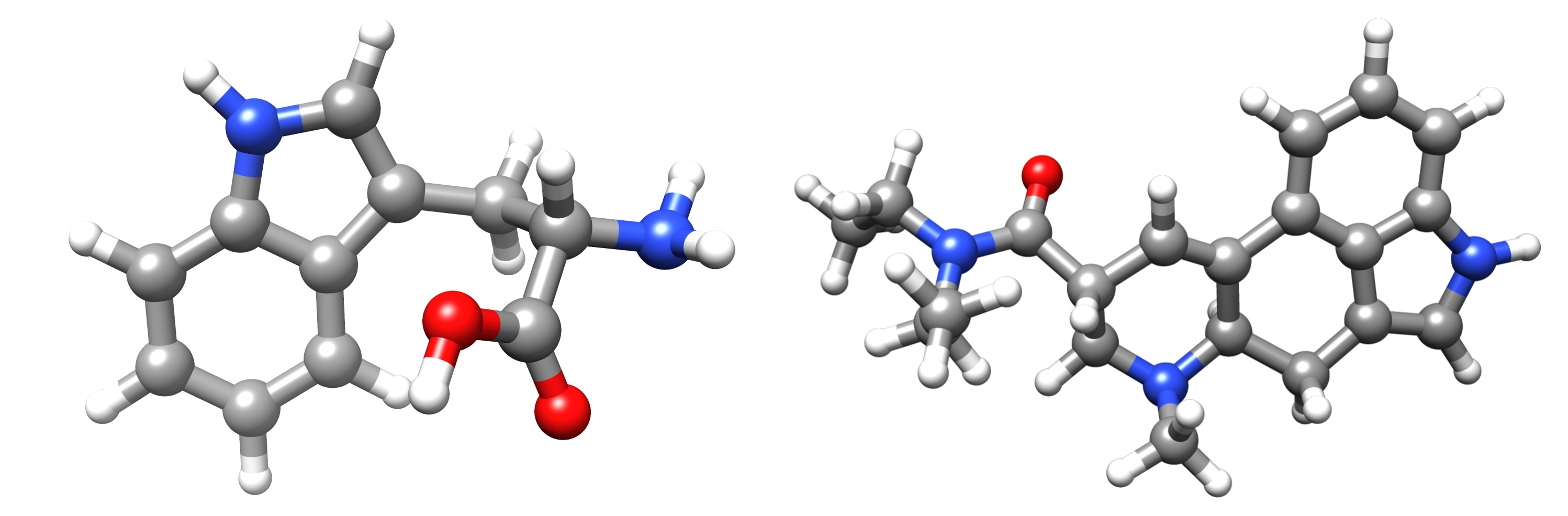}
    \caption{Tryptophan (left) and LSD (right)}
    \label{fig:tryptophan_lsd}
\end{figure}

\begin{table}
    \caption{CCSD/aug-cc-pVDZ excitation energies $(\omega)$ 
    and oscillator strengths $(f_\omega)$ for tryptophan.}
    \centering
    \begin{tabular}{c c c}
        \toprule
        & $\omega~[\si{\eV}]$ & $f_\omega$\\
        \midrule
        $S_1$ & 4.806 & 0.032 \\
        $S_2$ & 4.821 & 0.001 \\
        $S_3$ & 4.972 & 0.088 \\
        $S_4$ & 5.364 & 0.001 \\
        \bottomrule
    \end{tabular}
    \label{tab:tryptophan_energies}
\end{table}

\begin{table}
    \caption{Total calculation times for CCSD/aug-cc-pVDZ ground state ($t_{\mrm{gs}}$),
    excitation energy ($t_{\omega}$),
    and oscillator strength calculations ($t_{f_\omega}$) for tryptophan.
    $n_{\mrm{calls}}^{\mrm{gs}}$ is the number of calculations of the residual vector 
    for the ground state.
    $n_{\mrm{calls}}^{\mrm{R}}$ and $n_{\mrm{calls}}^{\mrm{L}}$ 
    are the number of calls to the Jacobian and Jacobian transpose transformations,
    respectively. The calculations were performed on an 
    Intel Xeon E5-2699 v4 using 44 threads and 1.5 TB shared memory.}
    \centering
    \begin{tabular}{l c c c c c c c}
        \toprule
         & $t_{\mrm{gs}}~[\si{\minute}]$
         & $t_{\omega}~[\si{\hour}]$
         & $t_{f_{\omega}}~[\si{\hour}]$
         & $n_{\mrm{calls}}^{\mrm{gs}}$
         & $n_{\mrm{calls}}^{\mrm{R}}$ 
         & $n_{\mrm{calls}}^{\mrm{L}}$ \\
        \midrule
        Dalton 2018 & 1409 & 84 & -- & 18 & 88 &  -- \\
        \eTone~     &  201 & 24 & 53 & 16 & 79 & 81 \\
        QChem 5.0   &  196 & 15 & 20 & 18 & 90 & 27 \\
         \bottomrule
    \end{tabular}
    \label{tab:tryptophan_timings}
\end{table}

\begin{table}
    \centering
    \caption{The FC-CCSD/aug-cc-pVDZ correlation energy $(E_{\text{corr}})$ and lowest excitation energy $(\omega)$ of LSD. A set of  decomposition thresholds $(\tau)$ for the Cholesky decomposition of the electron repulsion integral matrix were used. Both the ground and excited state equations are converged to within a residual threshold of $10^{-6}$.
    Deviations in the correlation and excitation energies ($\Delta E_{\text{corr}}$ and $\Delta \omega$), are relative to $\tau=10^{-8}$}
    \begin{tabular}{c c r c r}
        \toprule
        $\tau$ & $E_{\text{corr}}~[\si{\hartree}]$ & $\Delta E_{\text{corr}}~[\si{\hartree}]$ & $\omega~[\si{\hartree}]$ & $\Delta \omega~[\si{\hartree}]$\\
        \midrule
         $10^{-2}$ & -3.6496733 &  $2.3\cdot10^{-2}$  & 0.1657343 &  $7.1\cdot10^{-4}$\\
         $10^{-3}$ & -3.6720218 &  $2.3\cdot10^{-4}$  & 0.1650370 &  $7.7\cdot10^{-6}$\\
         $10^{-4}$ & -3.6723421 &  $-9.2\cdot10^{-5}$ & 0.1650279 & $-1.4\cdot10^{-6}$\\
         $10^{-6}$ & -3.6722542 &  $-3.6\cdot10^{-6}$ & 0.1650294 &  $1.1\cdot10^{-7}$\\
         $10^{-8}$ & -3.6722506 &  \multicolumn{1}{c}{--} & 0.1650293 & \multicolumn{1}{c}{--}\\
         \bottomrule
    \end{tabular}
    \label{tab:lsd_ccsd}
\end{table}

At the CCSD level of theory, 
we report calculations for the amino acid tryptophan\citep{tryptophanPUBCHEM} (chemical formula C$_{11}$H$_{12}$N$_{2}$O$_{2}$) 
and excitation energies for the psychoactive agent lysergic acid diethylamide (LSD)\citep{lsd} (chemical formula C$_{20}$H$_{25}$N$_{3}$O$_{1}$). 
Tryptophan and LSD are depicted in Figure \ref{fig:tryptophan_lsd}. 

For tryptophan, 
we have determined the four lowest excitation energies and the corresponding oscillator strengths at the CCSD/aug-cc-pVDZ level of theory ($n_{\mrm{MO}} = 453$). 
Energies and oscillator strengths are reported in Table \ref{tab:tryptophan_energies}. 
Timings for \eTone, Dalton 2018, and QChem 5.0 
are given in Table \ref{tab:tryptophan_timings}.
Thresholds in \eT~were set to target an energy convergence of $10^{-6}$: 
the residuals were converged to $10^{-6}$ for the ground state 
and $10^{-3}$ for the excited states (assuming quadratic errors for the energy).
In QChem, 
thresholds for ground and excited states were set to $10^{-6}$.
We report the total wall time for each calculation. The excited state timing includes the time to converge ground state and excited state equations.
The oscillator strength timing also includes the time to solve the multiplier and the left excited state equations.
\eT~and QChem are equally efficient for the CCSD ground state,
while Dalton is considerably slower.
For the CCSD excited state calculation, QChem reduced the wall time by a factor of 1.6 compared to \eT~and a factor of 5.6 compared to Dalton. 
For the oscillator strength calculations,
QChem reduced the wall time by a factor of 2.7 compared to \eT. 
The superior performance of QChem for oscillator strengths is primarily due to an efficient starting guess for the left excitation vectors: 
only 27 linear transformations are needed to converge all four roots.

We have performed FC-CCSD/aug-cc-pVDZ calculations on LSD ($n_{\mrm{MO}} = 777$, $n_{\mrm{frozen}} = 24$). To demonstrate the effect of
integral approximation through Cholesky decomposition, we consider a range of decomposition thresholds.
The correlation energy and the lowest excitation energy are given in Table \ref{tab:lsd_ccsd}.
Both ground and excited state 
residual thresholds are $10^{-6}$.
With a decomposition threshold of $10^{-2}$,
the error in the excitation energy ($\Delta\omega$)
is less than $10^{-3}\si{\hartree}$, well
within the expected accuracy of 
FC-CCSD.\citep{Christiansen1996,kannar2014benchmarking,kannar2016accuracy}

\begin{figure}
    \centering
    \includegraphics[width=0.4\textwidth]{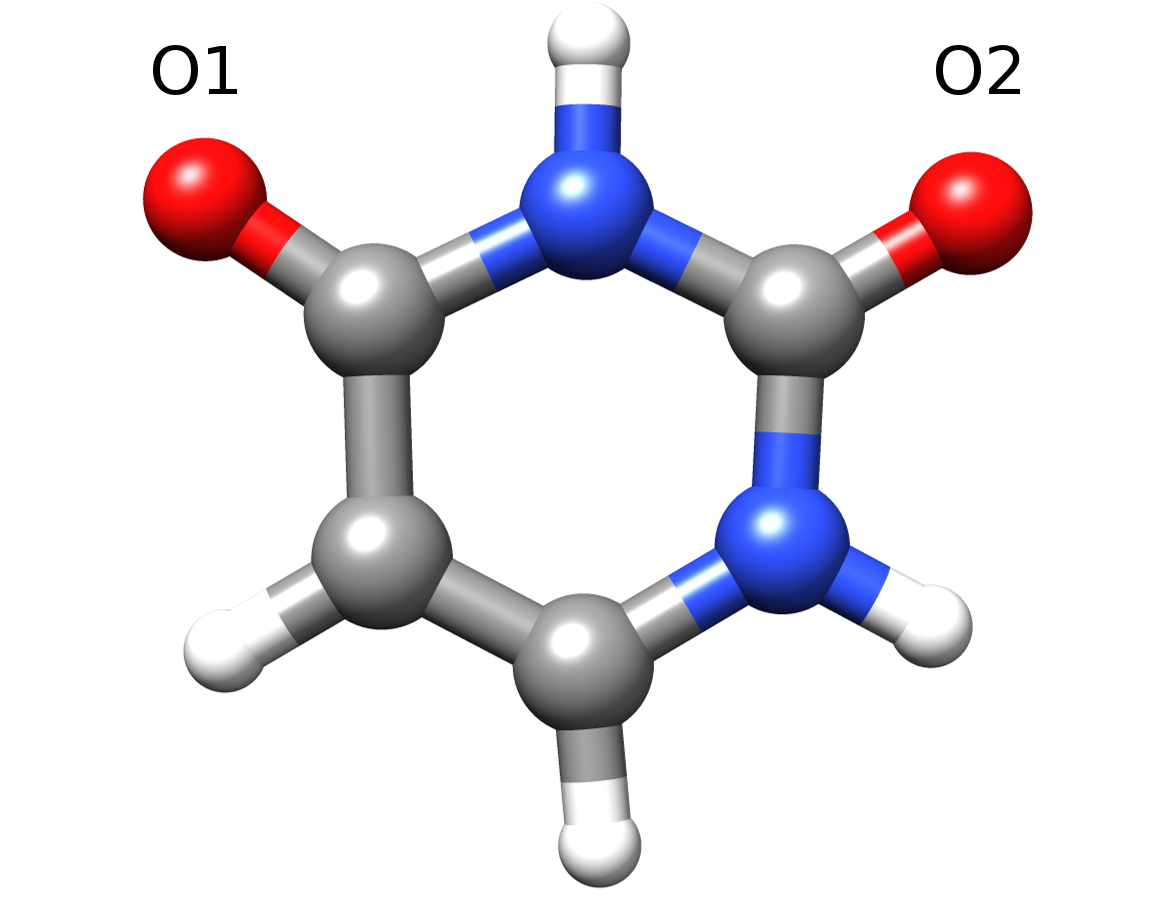}
    \caption{Uracil with labels on the oxygens}
    \label{fig:uracil}
\end{figure}

\begin{table}
    \centering
    \caption{
    Timing comparisons of \eTone~and Dalton 2018 for the lowest CC3/aug-cc-pVDZ excitation energy ($\omega$) of glycine. 
    The calculations were performed on a node with two Intel Xeon Gold 6138 processors, 
    using 40 threads and 320 GB shared memory.}
    \label{tab:glycine}
    \begin{tabular}{c c c c}
    \toprule
    $\omega$ & \eT~& Dalton (new) & Dalton (old) \\
    \midrule
    $\SI{5.879}{\eV}$ & $\SI{33}{\minute}$ & $\SI{73}{\minute}$ & $\SI{1279}{\minute}$\\
    \bottomrule
    \end{tabular}
\end{table}

\begin{table}
    \centering
    \caption{CC3 valence and core (oxygen edge) excitation energies $(\omega)$, 
    and EOM oscillator strengths $(f_{\omega})$ for uracil. 
    Valence excitations were calculated with the aug-cc-pVTZ basis on all atoms 
    and the frozen core approximation. 
    Core excitations were calculated using the CVS approximation
    with the aug-cc-pCVTZ basis on the oxygen atom being excited 
    and the aug-cc-pVDZ basis on the remaining atoms.}
    \label{uracil_results}
    \begin{tabular}{c c c c c}
    \toprule
     & \multicolumn{2}{c}{CCSD}& \multicolumn{2}{c}{CC3} \\
     & $\omega~[\si{\eV}]$ & $f_{\omega}$ 
     & $\omega~[\si{\eV}]$ & $f_{\omega}$\\
    \midrule
    Valence & 
    $5.08$ & $2.24\cdot10^{-8}$ &
    $4.81$ & $2.23\cdot10^{-6}$\\
    \midrule
    \multirow{2}{*}{core O1} 
    & $536.04$ & $3.35\cdot10^{-2}$ 
    & $533.64$ & $1.95\cdot10^{-2}$ \\
    & $539.60$ & $3.23\cdot10^{-4}$ 
    & $535.66$ & $2.24\cdot10^{-4}$ \\
    \midrule
    \multirow{2}{*}{core O2} 
    & $536.98$ & $3.13\cdot10^{-2}$ 
    & $534.64$ & $1.32\cdot10^{-2}$ \\
    & $539.44$ & $1.47\cdot10^{-4}$ 
    & $535.75$ & $1.34\cdot10^{-4}$ \\
    \bottomrule
    \end{tabular}
\end{table}

\begin{table}
   \centering
   \caption{Total wall times for CC3 on uracil.
   The valence calculation was performed on a node with two Intel Xeon Gold 6138 processors, 
   using 40 threads and 320 GB shared memory.
   The CVS calculations were performed on a node with two Intel Xeon Gold 6138 processors, 
   using 40 threads and 150 GB shared memory.
   n$_o$ and n$_v$ are the number of occupied and virtual orbitals,
   respectively.}
   \label{cc3_Uracil_times}
   \begin{tabular}{lcccc}
   \toprule
   Calculation & basis set & $t~[\si{\hour}]$ & n$_o$ & n$_v$ \\
   \midrule
   Valence excitation & aug-cc-pVTZ    & 147 & 21 & 431 \\
   CVS O1             & aug-cc-pV(CT)Z &  36 & 29 & 227 \\
   CVS O2             & aug-cc-pV(CT)Z &  38 & 29 & 227 \\
   \bottomrule
\end{tabular}
\end{table}

\begin{table}
   \centering
   \caption{Average wall time per function call for both
   CC3 core excitation calculations on uracil. 
   $n_\mrm{calls}$ is the total number of routine calls in the two calculations.}
   \label{cc3_Uracil_CVS}
   \begin{tabular}{lcc}
   \toprule
   Contributions              &  $t~[\si{\minute}]$ & $n_{\mrm{calls}}$\\
   \midrule
   Ground state amplitudes         & 14 &  28\\
   Ground state multipliers        & 23 &  30\\
   Right excited states            &  4 & 195\\
   Left excited states             &  7 & 244\\
   \bottomrule
\end{tabular}
\end{table}

The CC3 model can be used to obtain highly accurate excitation energies. 
However, 
an iterative cost that scales as $\mathcal{O}(n_\mrm{v}^4n_\mrm{o}^3)$ 
severely limits system size. 
To the best of our knowledge, \eT~includes
the fastest available implementation of CC3.
In Table \ref{tab:glycine}, 
we compare \eT~timings to the new\citep{myhre2016multilevel}
and efficient CC3 implementation,
as well as the old\citep{CC3imp} implementation,
in Dalton 2018.\citep{dalton} 
The example system is glycine (chemical formula C$_{2}$H$_{5}$NO$_{2}$) 
with a geometry optimized 
at the CCSD(T)/aug-cc-pVDZ level using CFOUR.\citep{stanton2010cfour} 
We have calculated ground and excited state energies and
\eT~is twice as fast as the new implementation in Dalton.

We have calculated valence and core excitation energies 
and EOM oscillator strengths for the nucleobase uracil 
(chemical formula C$_4$H$_4$N$_2$O$_2$, 
see Figure \ref{fig:uracil}).
The geometry was optimized
at the CCSD(T)/aug-cc-pVDZ level using CFOUR.\citep{stanton2010cfour}
One valence excitation energy was calculated
at the FC-CCSD/aug-cc-pVTZ 
and FC-CC3/aug-cc-pVTZ levels of theory ($n_\mrm{MO} = 452$). 
Two core excited states were calculated for each of the oxygen atoms
(O1 and O2, see Figure \ref{fig:uracil}) 
at the CCSD and CC3 levels. The
aug-cc-pCVTZ basis was used on the oxygen being excited 
and aug-cc-pVDZ on the remaining atoms ($n_\mrm{MO} = 256$). 
The results are given in Table \ref{uracil_results}.
Total timings for the uracil calculations are presented in Table \ref{cc3_Uracil_times}. In Table \ref{cc3_Uracil_CVS}, we present averaged timings from the CVS calculations. 
They clearly demonstrate the reduced computational cost 
of the CVS implementation for CC3.
The ground state calculation was about four times more expensive per iteration
than the right excited state.  
Without the CVS approximation,
the computational cost of the excited states scale as $4n_\mrm{v}^4n_\mrm{o}^3$
per iteration while the ground state scales as $2n_\mrm{v}^4n_\mrm{o}^3$.
Using CVS,
the excited state scaling is reduced to $4n_\mrm{v}^4n_\mrm{o}^2$.

\subsection{Cholesky decomposition}

\begin{figure}
    \centering
    \includegraphics[width=\linewidth]{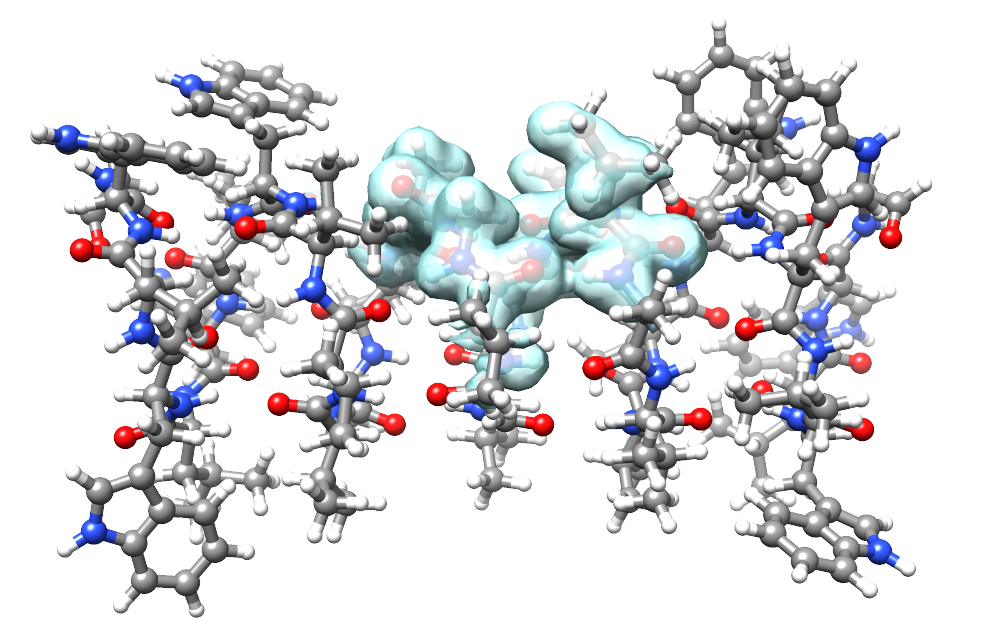}
    \caption{Gramicidin A.
    The active MLHF/cc-pVDZ density is shown.}
    \label{fig:gramicidin}
\end{figure}

\begin{table}[htbp]
    \centering
    \caption{Cholesky decomposition wall times ($t$) for gramicidin. 
    $\tau$ is the decomposition threshold and
    $n_J$ the number of Cholesky vectors.
    For reference, 
    we include the time 
    $(t^\mathrm{HF}_\mathrm{it})$ 
    for one full Hartree-Fock iteration. 
    All calculations were performed on an 
    Intel Xeon E5-2699 v4 using 44 threads and 1.5 TB shared memory.}
    \label{tab:gramicidin_cholesky}
    \begin{tabular}{l c c c c c}
        \toprule
        Basis & 
        $n_\mathrm{AO}$ & 
        $\tau$ & 
        $n_J$ & 
        $t~[\si{\minute}]$ & 
        $t^\mathrm{HF}_\mathrm{it}~[\si{\minute}]$ \\
        \midrule
        \multirow{4}{*}{cc-pVDZ} & \multirow{4}{*}{5188} 
          & $10^{-2}$ & 11574 &   3 & \multirow{4}{*}{35} \\
        & & $10^{-3}$ & 16368 &   6 &  \\
        & & $10^{-4}$ & 24652 &  12 &  \\
        & & $10^{-8}$ & 75446 & 125 &  \\
         \midrule
        \multirow{4}{*}{aug-cc-pVDZ} & \multirow{4}{*}{8740} 
          & $10^{-2}$ & 12813 &  8   & \multirow{4}{*}{1191} \\
        & & $10^{-3}$ & 18587 &  27  & \\
        & & $10^{-4}$ & 29818 &  61  &\\
        & & $10^{-8}$ & 90656 & 645  & \\
         \bottomrule 
    \end{tabular}
\end{table}

We have determined the Cholesky basis for the transmembrane ion channel gramicidin A
(chemical formula C$_{198}$H$_{276}$N$_{40}$O$_{34}$, see Figure \ref{fig:gramicidin}). The geometry is taken 
from the supporting information of Ref.~\citenum{gramicidinGeo}.
Decomposition times are given in Table \ref{tab:gramicidin_cholesky} for the cc-pVDZ and aug-cc-pVDZ basis sets and a range of decomposition thresholds. 
These are compared to the time of one HF 
iteration.
Except when using cc-pVDZ with the tightest threshold,
the decomposition time is small or negligible
compared to one Fock matrix construction. 

\subsection{Hartree-Fock}

\begin{figure}
    \centering
    \includegraphics[width=\linewidth]{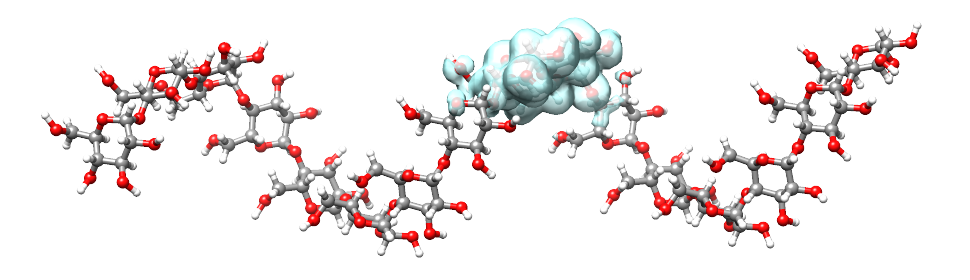}
    \caption{Amylose chain of 16 glucose units.
    The active MLHF/cc-pVDZ density is shown.}
    \label{fig:amylose}
\end{figure}

\begin{table*}[htbp]
    \centering
    \caption{Hartree-Fock/cc-pVDZ calculations on amylose and gramicidin. 
    The total wall time is denoted by $t$ and $\tau_{\mrm{SCF}}$ is the Hartree-Fock convergence threshold.
    We present timings for \eT~and QChem, 
    along with the computed Hartree-Fock energies $(E)$
    and absolute energy differences $(|\Delta E|)$
    with respect to the calculation with the tightest threshold. 
    Calculations were performed on two Intel Xeon E5-2699 v4 processors, 
    using 44 threads and 1.5 TB shared memory.}
    \label{tab:amylose_gramicidin_hf}
    \begin{tabular}{c l c r c c l c c c}
        \toprule
         &  \multicolumn{4}{c}{\eT} & & \multicolumn{4}{c}{QChem}  \\ 
         \cmidrule{2-5} 
         \cmidrule{7-10}
           & $\tau_\mathrm{SCF}$ 
           & $E~[\si{\hartree}]$ 
           & $|\Delta E|~[\si{\hartree}]$
           & $t~[\si{\minute}]$ 
           & 
           & $\tau_\mathrm{SCF}$ 
           & $E~[\si{\hartree}]$ 
           & $|\Delta E|~[\si{\hartree}]$
           & $t~[\si{\minute}]$ \\
        \midrule 
        \multirow{6}{*}{amylose}
        &  $10^{-3}$ & $-9792.08512990$ &  $4\cdot10^{-5}$ &  21 
        & 
        &  $10^{-5}$ & $-9792.08535039$ &  $2\cdot10^{-4}$ &   9
        \\
        &  $10^{-4}$ & $-9792.08517833$ &  $5\cdot10^{-6}$ &  31 
        & 
        &  $10^{-6}$ & $-9792.08518084$ &  $7\cdot10^{-6}$ &  14
        \\
        &  $10^{-5}$ & $-9792.08517442$ &  $7\cdot10^{-7}$ &  42
        & 
        &  $10^{-7}$ & $-9792.08517119$ &  $2\cdot10^{-6}$ &  19
        \\ 
        &  $10^{-6}$ & $-9792.08517377$ &  $1\cdot10^{-8}$ &  60 
        & 
        &  $10^{-8}$ & $-9792.08517323$ &  $4\cdot10^{-7}$ &  26
        \\
        &  $10^{-7}$ & $-9792.08517376$ & $<1\cdot10^{-8}$ &  78 
        & 
        &  $10^{-9}$ & $-9792.08517361$ &  $3\cdot10^{-8}$ &  33
        \\
        & $10^{-10}$ & $-9792.08517376$ & \multicolumn{1}{c}{--}    & 153
        & 
        & $10^{-10}$ & $-9792.08517358$ & \multicolumn{1}{c}{--}    &  46 
        \\
        \midrule 
        \multirow{3}{*}{gramicidin} 
        & $10^{-4}$ & $-12383.45883254$ & $4\cdot10^{-6}$ & 130
        & 
        & $10^{-6}$ & $-12383.45882513$ & $1\cdot10^{-5}$ &  50
        \\
        & $10^{-5}$ & $-12383.45883634$ & $7\cdot10^{-8}$ & 198
        & 
        & $10^{-7}$ & $-12383.45882710$ & $1\cdot10^{-5}$ &  77
        \\
        & $10^{-6}$ & $-12383.45883627$ & \multicolumn{1}{c}{--}   & 280
        & 
        & $10^{-8}$ & $-12383.45883677$ & \multicolumn{1}{c}{--}   & 111
        \\
        \bottomrule
    \end{tabular}
\end{table*}

Systems with several hundred atoms are easily modelled in \eT~using Hartree-Fock.
In Table \ref{tab:amylose_gramicidin_hf}, 
we present wall times for calculations on
gramicidin A (see Figure \ref{fig:gramicidin}) and an amylose chain with 16 glucose units (chemical formula C$_{96}$H$_{162}$O$_{81}$, see Figure \ref{fig:amylose}).
The amylose geometry is taken from Sæther \emph{et al.}\citep{saether2017density}
We compare results and timings from \eTone~and QChem 5.0.\citep{shao2015qchem}
This comparison is 
complicated because the 
accuracy depends on several thresholds apart from the gradient and energy thresholds, 
e.g.,
screening thresholds and integral accuracy.
We therefore list energies and absolute energy differences along with the timings in Table \ref{tab:amylose_gramicidin_hf}.
QChem 5.0 outperforms 
\eT~by about factor of two.
The energies converge to slightly different results in the two programs. 
In the case of amylose, 
we find a $2 \cdot {10^{-7}}~\si{\hartree}$ energy difference 
using the tightest thresholds ($\tau_\mathrm{SCF} = 10^{-10}$). 
We are able to reproduce the \eT~results using tight thresholds in LSDalton 2018.\citep{dalton}

\subsection{Multilevel and multiscale methods}

\begin{table*}[htbp]
    \centering
    \caption{
    The two lowest excitation energies 
    ($\omega_1$ and $\omega_2$) of $\text{SO}_2$ with 21 water molecules, 
    calculated with full, and reduced space FC-CC2 and FC-MLCC2, using HF and MLHF reference wave functions. 
    Deviation from full FC-CC2 ($\Delta\omega_i = \omega_i - \omega_i^{\text{FC-CC2}}$) are given.
    We also list the number of  
    occupied ($n_\mrm{o}$) and virtual ($n_\mrm{v}$) orbitals 
    treated at the different levels of theory.
    There are a total of $121$ occupied orbitals and $813$ virtual orbitals in the system.}
    \label{tab:so2}
    \begin{tabular}{l c c c c c c c c c c}
    \toprule
    \multirow{2}{*}{Calculation} & \multicolumn{2}{c}{HF} &
    \multicolumn{2}{c}{CCS} & \multicolumn{2}{c}{CC2} &\multirow{2}{*}{$\omega_1~[\si{\eV}]$}
    &\multirow{2}{*}{$\Delta\omega_1~[\si{\eV}]$}
    &\multirow{2}{*}{$\omega_2~[\si{\eV}]$}
    & \multirow{2}{*}{$\Delta\omega_1~[\si{\eV}]$}\\
    & $n_\mrm{o}$ & $n_\mrm{v}$ & $n_\mrm{o}$ & $n_\mrm{v}$ & $n_\mrm{o}$ & $n_\mrm{v}$ & & \\
        \midrule
    FC-CC2 
    & 121 & 813 
    & -- & -- 
    & 93 & 813 
    & 3.11 & -- & 3.39 & -- \\ 
    FC-CC2-in-HF 
    & 121 & 813 
    & -- & -- 
    & 40 & 266 
    & 3.14 & 0.03 & 3.43 & 0.04 \\
    FC-CC2-in-MLHF 
    & 75 & 426 
    & -- & -- 
    & 40 & 266 
    & 3.16 & 0.05 & 3.44 & 0.05 \\ 
    FC-MLCC2 
    & 121 & 813 
    & 93 & 813 
    & 14 & 67 
    & 3.18 & 0.07 & 3.45 & 0.06 \\
    FC-MLCC2-in-HF 
    & 121 & 813 
    & 40 & 266 
    & 14 & 66 
    & 3.18 & 0.07 & 3.45 & 0.06 \\
    FC-MLCC2-in-MLHF 
    & 75 & 426 
    & 40 & 266 
    & 15 & 66 
    & 3.20 & 0.09 & 3.47 & 0.08 \\
    \bottomrule
    \end{tabular}
\end{table*}

\begin{figure}
    \centering
    \includegraphics[width=\linewidth]{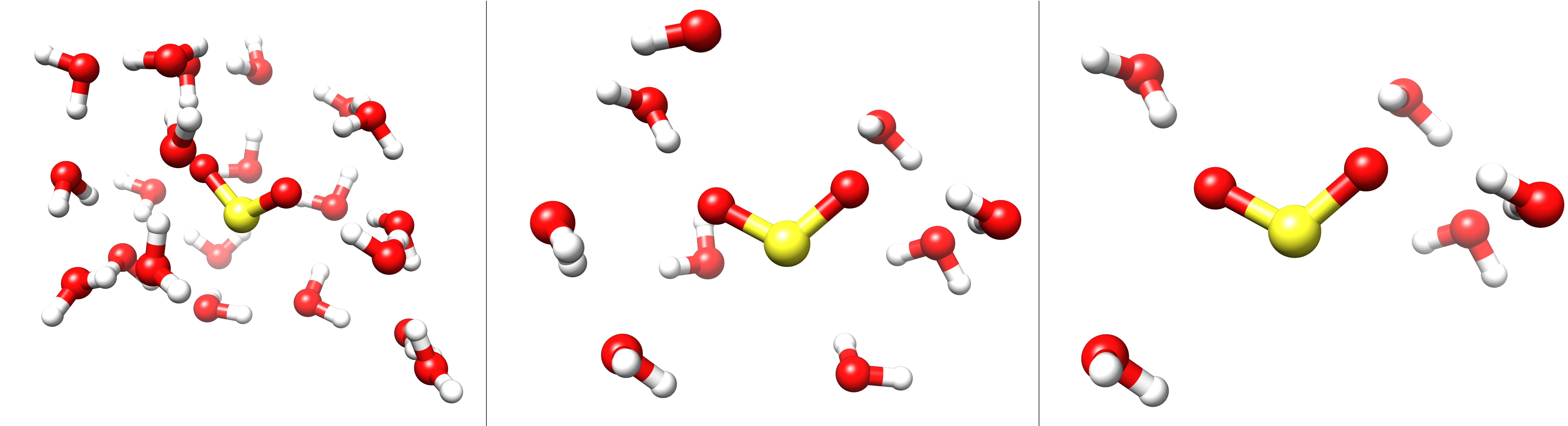}
    \caption{$\text{SO}_2$ and water. On the left, $\text{SO}_2$ and 21 water molecules. 
    In the middle, $\text{SO}_2$ and nine water molecules; 
    these are the HF active atoms in the MLHF calculations.
    On the right, $\text{SO}_2$ and five water molecules; 
    these are the CC active atoms.
    In the MLCC2 calculations only $\text{SO}_2$ is treated at the CC2 level of theory.}
    \label{fig:so2}
\end{figure}

To demonstrate the efficacy of multilevel methods for excitation energies, 
we consider a system of sulfur dioxide with 21 water molecules, 
(see Figure \ref{fig:so2}). 
In Table \ref{tab:so2}, 
we present different flavours of multilevel calculations 
to approximate the two lowest FC-CC2 excitation energies for this system. 
Three sets of active atoms are defined.
The first set contains the sulfur dioxide and nine water molecules; 
these atoms determine the active orbitals of the MLHF calculation. 
The second set contains the sulfur dioxide and five water molecules; 
these atoms determine the reduced space coupled cluster calculations. 
The third set contains only sulfur dioxide
and determines the CC2 active space in the MLCC2 calculations. 
The reduced space FC-CC2 calculations are denoted FC-CC2-in-HF and FC-CC2-in-MLHF 
and similarly for the reduced space FC-MLCC2 calculations.
Orbital spaces are partitioned using Cholesky occupied orbitals and PAOs for the virtual orbitals. 
In all calculations, 
the deviation with respect to full FC-CC2 is within the expected error of CC2, 
which is about $\SI{0.1}{\eV}$.\cite{kannar2014benchmarking,kannar2016accuracy}

\begin{table}[htbp]
    \centering
    \caption{Multilevel Hartree-Fock wall times for amylose and gramicidin. 
    $t_\mathrm{it}$ is the wall time to construct the Fock matrix. 
    For the calculations with (aug)-cc-pVDZ,
    aug-cc-pVDZ is used on the active atoms and cc-pVDZ for the rest.
    The total number of AOs and the active MOs are labelled $n_\mathrm{AO}$ and
    $n_\mathrm{MO}^\mathrm{active}$. 
    Thresholds for Coulomb and exchange are set to $10^{-12}$ and $10^{-10}$, 
    respectively, 
    and the integral cutoff to $10^{-12}$.
    Calculations were performed on two Intel Xeon E5-2699 v4 processors, 
    using 44 threads and 1.5 TB shared memory.}
    \label{tab:amylose_gramicidin_mlhf}
    \begin{tabular}{l l c c c c}
        \toprule
        &       & \multicolumn{2}{c}{HF} & \multicolumn{2}{c}{MLHF}  \\
        & Basis & $n_\mathrm{AO}$ & $t_\mathrm{it}~[\si{\minute}]$ & 
        $n_\mathrm{MO}^\mathrm{active}$ & $t_\mathrm{it}~[\si{\minute}]$ \\
        \midrule 
        \multirow{2}{*}{amylose} 
        &       cc-pVDZ & 3288 &  8 & 335 & 1 \\
        & (aug)-cc-pVDZ & 3480 & 11 & 552 & 4 \\
        \midrule  
        \multirow{2}{*}{gramicidin} 
        &       cc-pVDZ & 5188 & 35 & 546 & 11 \\
        & (aug)-cc-pVDZ & 5506 & 69 & 942 & 50 \\
        \bottomrule 
    \end{tabular}
\end{table}

In order to assess the performance of the MLHF implementation, 
we  compare full HF and MLHF for gramicidin A and amylose.
Active electron densities from the MLHF calculations are
shown in Figures \ref{fig:gramicidin} and \ref{fig:amylose}.
The plots were generated using UCSF Chimera.\citep{Pettersen2004Chimera} 
Cholesky orbitals were used to partition
the occupied space 
and PAOs 
were used for the virtual space.
We present timings in Table \ref{tab:amylose_gramicidin_mlhf}.
For amylose, 
the iteration times are reduced significantly with MLHF: 
by a factor of eight when cc-pVDZ is used on all atoms
and a factor of three when aug-cc-pVDZ is used on the active atoms.
In contrast, only a factor of three was reported by Sæther \emph{et al.}\citep{saether2017density} in the cc-pVDZ case.
The iteration time is also reduced by a factor of eight 
for amylose/cc-pVDZ ($t_\mathrm{iteration} = 1$ m,
$n_\mathrm{MO}^\mathrm{active} = 318$) when using Cholesky virtuals 
(as in Ref.\citenum{saether2017density}) instead of PAOs.
The savings for amylose reflect the small active region 
as well as the linear structure of the chain. 
Savings are less significant for the gramicidin system, 
where 
the MLHF iteration time is a third of the HF iteration time
for cc-pVDZ, 
but only about two thirds 
when the active atoms are described using aug-cc-pVDZ. 
The smaller savings reflect the relatively large active region 
and the more compact shape of the gramicidin system.

\begin{figure}
\centering
\includegraphics[width=\linewidth]{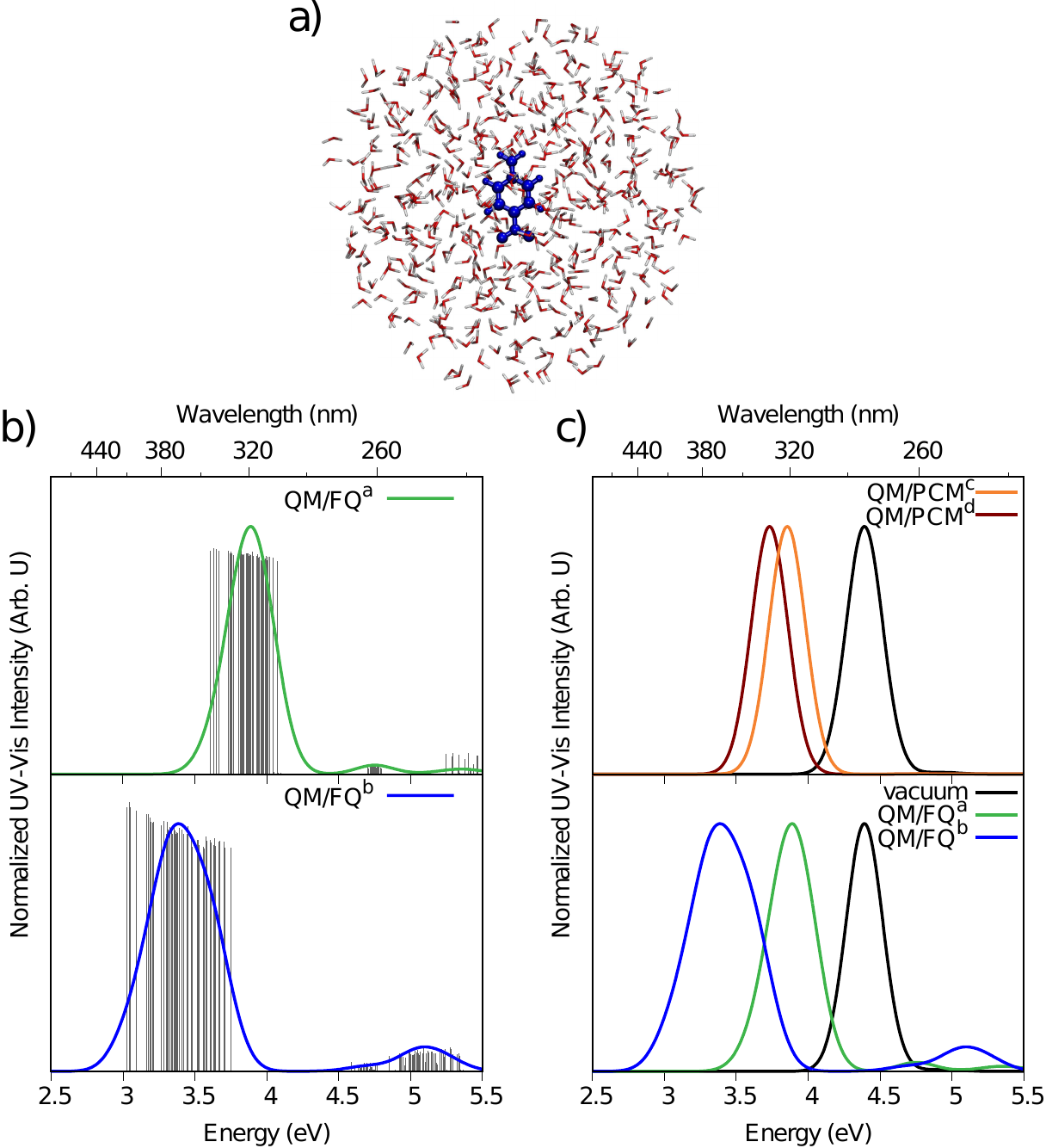}
\caption{\textbf{a)} Schematic representation of a random snapshot of PNA in aqueous solution.
\textbf{b)} and \textbf{c)}
UV/Vis spectra of PNA calculated at the CC2/aug-cc-pVDZ level of theory 
with an aqueous solution described at the PCM or FQ level of theory.
\textbf{b)} QM/FQ raw data (sticks) together 
with their Gaussian convolution (FWHM = 0.3 eV). 
\textbf{c)} QM/PCM (top) and QM/FQ (bottom) spectra in aqueous solution. 
A gas phase CC2/aug-cc-pVDZ reference spectrum is also reported (black). 
For QM/FQ$^\text{a}$, 
the FQ parametrization is from Ref. \citenum{rick1994dynamical} and for QM/FQ$^\text{b}$, 
the parametrization is from Ref. \citenum{giovannini2019epr}.
In QM/FQ$^\text{c}$, 
the PCM cavity is constructed using the 
UFF radii,\cite{uffradii} 
and in QM/FQ$^\text{d}$ it is constructed using the 
Bondi radii.\cite{bondiradii}
}
\label{fig:qmmm}
\end{figure}

\begin{table}[htbp]
\caption{The first vertical excitation energy of PNA in vacuum ($\omega_\mrm{v}$)
and in aqueous solution ($\omega_\mrm{s}$), 
as well as water-to-vacuum solvatochromatic shifts $(\Delta\omega)$.
PNA is treated at the CC2/aug-cc-pVDZ level of theory,
and the solution is described with PCM or FQ.
68\% confidence intervals for excitation energies are also reported for QM/FQ,
calculated as $\sigma/\sqrt{N}$, 
where $\sigma$ is the standard deviation 
and $N$ is the number of the snapshots used to obtain the average property. 
Experimental data are taken from Ref.~\citenum{kovalenko2000femtosecond}.}
\label{tab:qmmm}
\begin{threeparttable}
\begin{tabular}{l c l l}
\toprule
& $\omega_\mrm{v}$ $[\si{\eV}]$& $\omega_\mrm{s}$ $[\si{\eV}]$& $\Delta\omega$ $[\si{\eV}]$\\
\midrule
CC2  & 4.38 & \multicolumn{1}{c}{--} &\multicolumn{1}{c}{--} \\
CC2/FQ\tnote{a} & -- & 3.88$\pm$0.01 & 0.50$\pm$0.01 \\
CC2/FQ\tnote{b} & -- & 3.38$\pm$0.01 & 1.00$\pm$0.01 \\
CC2/PCM\tnote{c}& -- & 3.86 & 0.52 \\
CC2/PCM\tnote{d}& -- & 3.76 & 0.62 \\
Experiment\tnote{e} & 4.25 & 3.26 & 0.99\\
\bottomrule
\end{tabular}
\begin{tablenotes}
\scriptsize
    \item [a] {FQ parametrization taken from Ref. \citenum{rick1994dynamical}}
    \item [b] {FQ parametrization taken from Ref. \citenum{giovannini2019epr}}
    \item [c] {PCM cavity constructed by exploiting UFF radii.\cite{uffradii}}
    \item [d] {PCM cavity constructed by exploiting Bondi radii.\cite{bondiradii}}
    \item [e] {Ref. \citenum{kovalenko2000femtosecond}}
\end{tablenotes}
\end{threeparttable}
\end{table}

For systems in solution, 
electronic spectra can be calculated using QM/MM or QM/PCM.
Paranitroaniline (PNA) has an experimental vacuum-to-water solvatochromism of about $\SI{1}{\eV}$.\cite{kovalenko2000femtosecond}
For QM/PCM, we use two different atomic radii, 
UFF\citep{uffradii} (QM/PCM$^\text{c}$) and
Bondi\citep{bondiradii} (QM/PCM$^\text{d}$), and
the dielectric permittivity of water was set to $\varepsilon = 78.39$.
For QM/MM, 
64 snapshots were extracted from a classical molecular dynamics simulation.\citep{giovannini2019fqfmulinear}
See Figure \ref{fig:qmmm}a for an example structure.
The UV/Vis spectra were then computed by treating PNA 
at the CC2/aug-cc-pVDZ level and 
modelling the water using an FQ force field.
Here, 
we present results using two different FQ parametrizations:
QM/FQ$^\text{a}$ from Ref. \citenum{rick1994dynamical} and 
QM/FQ$^\text{b}$ from Ref. \citenum{giovannini2019epr}.
See the supplemental material for additional computational details.

The spectra calculated using QM/FQ are presented in Figure \ref{fig:qmmm}b.
Results for individual snapshots are presented as sticks 
together with their Gaussian convolution.
As can be seen from Figure \ref{fig:qmmm},
QM/FQ$^\mrm{b}$ results in a greater spread in the excitation energies.
This is probably due to the larger molecular dipole moments of
the water molecules in this 
parametrization.\cite{giovannini2019fqfmulinear,giovannini2019fqfmu}

In Figure \ref{fig:qmmm}c, 
convoluted spectra
calculated using QM/PCM$^\text{c}$ and QM/PCM$^\text{d}$ (top), 
and QM/FQ$^\text{a}$ and QM/FQ$^\text{b}$ (bottom), 
are presented with their vacuum counterparts.
The excitation energies are also given in Table \ref{tab:qmmm} 
together 
with experimental data from Ref.~\citenum{kovalenko2000femtosecond}.
For QM/FQ, we also report 68\% confidence intervals for the calculated excitation energies. 
QM/FQ$^\mrm{b}$ reproduces the experimental solvatochromism,
while the other approaches give errors of 40-50\%.

\subsection{Modelling spectroscopies} \label{sec:results:spectroscopic}

\begin{figure}
    \centering
    \includegraphics[width=\linewidth]{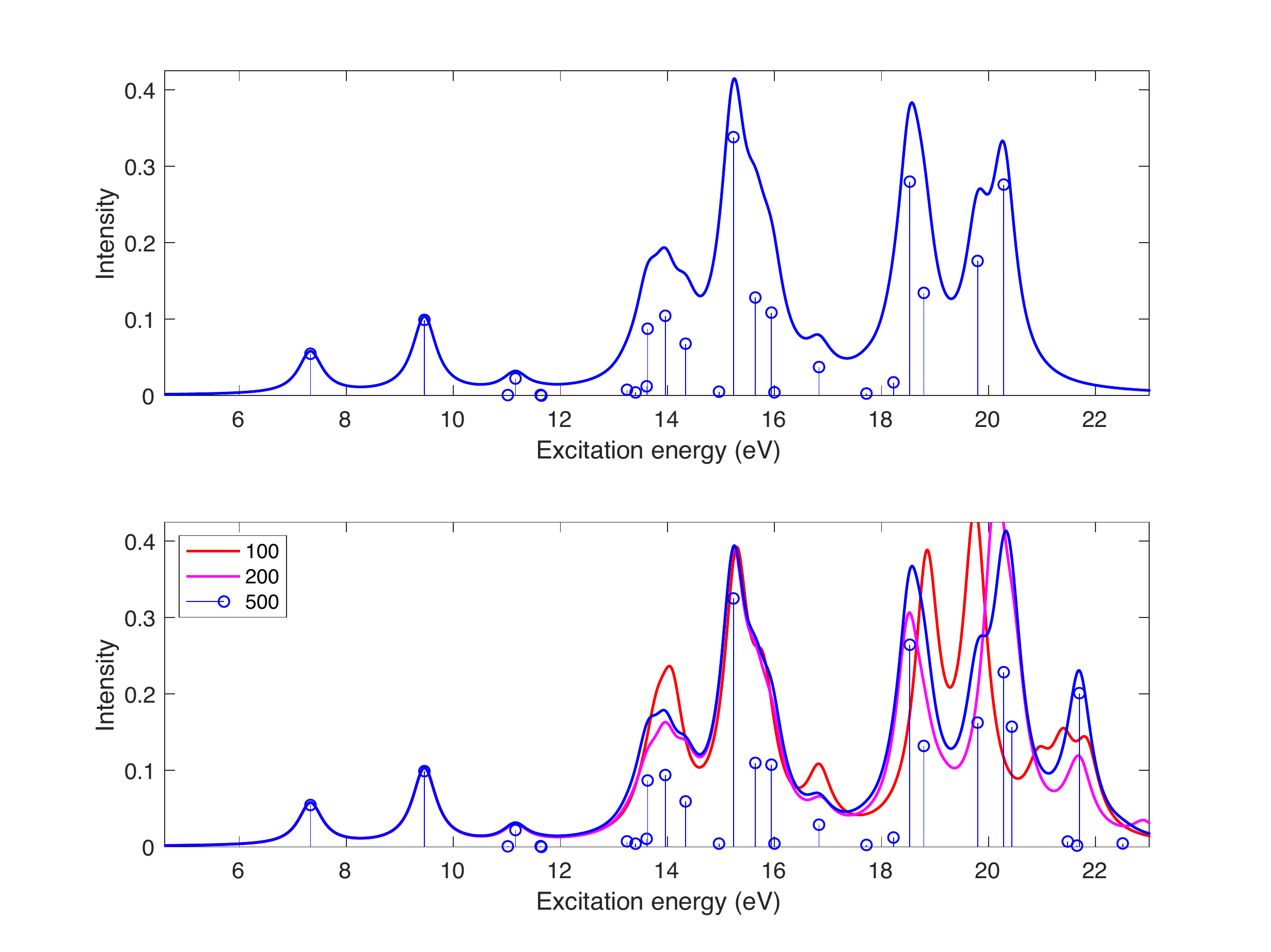}
    \caption{Water CCSD/aug-cc-pCVDZ UV/Vis absorption spectrum. Lorentzian broadening ($\SI{0.02}{\hartree}$~FWHM) has been applied to the stick spectra.
    The top plot shows the spectrum obtained using Davidson.
    The spectrum in the bottom plot is from Lanczos calculations 
    with chain lengths 
    $100$ (red), 
    $200$ (magenta), 
    and $500$ (blue).}
    \label{fig:valence_h2o}
\end{figure}

\begin{figure}
    \centering
    \includegraphics[width=\linewidth]{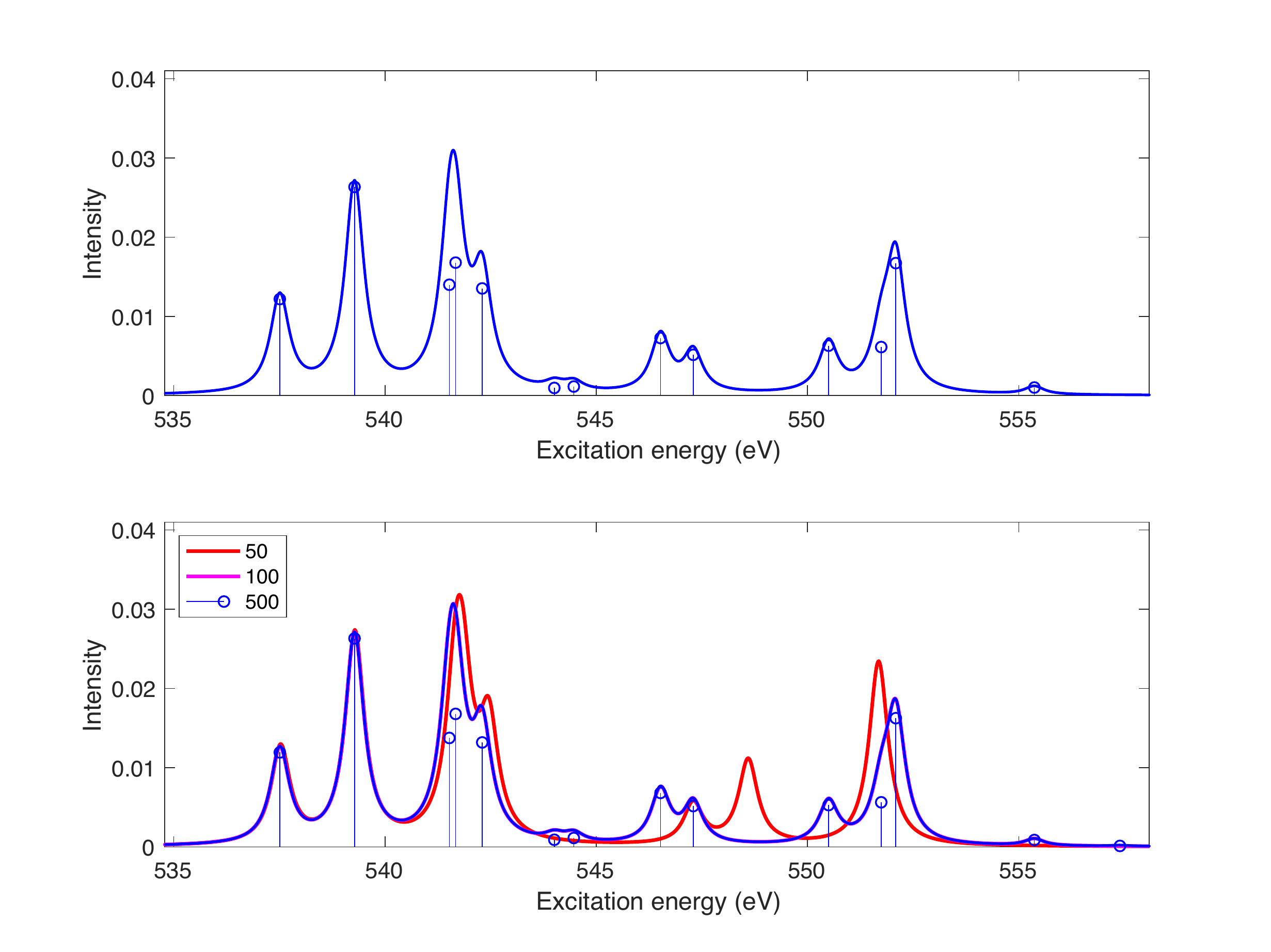}
    \caption{Water CCSD/CVS/aug-cc-pCVDZ X-ray absorption spectrum. Lorentzian broadening ($\SI{0.02}{\hartree}$~FWHM) has been applied to the stick spectra.
    The top plot shows the spectrum obtained using Davidson.
    The spectrum in the bottom plot is from Lanczos calculations 
    with chain lengths  
    $50$ (red), 
    $100$ (magenta), 
    and $500$ (blue).}
    \label{fig:cvs_h2o}
\end{figure}

\begin{figure}
    \centering
    \includegraphics[width=\linewidth]{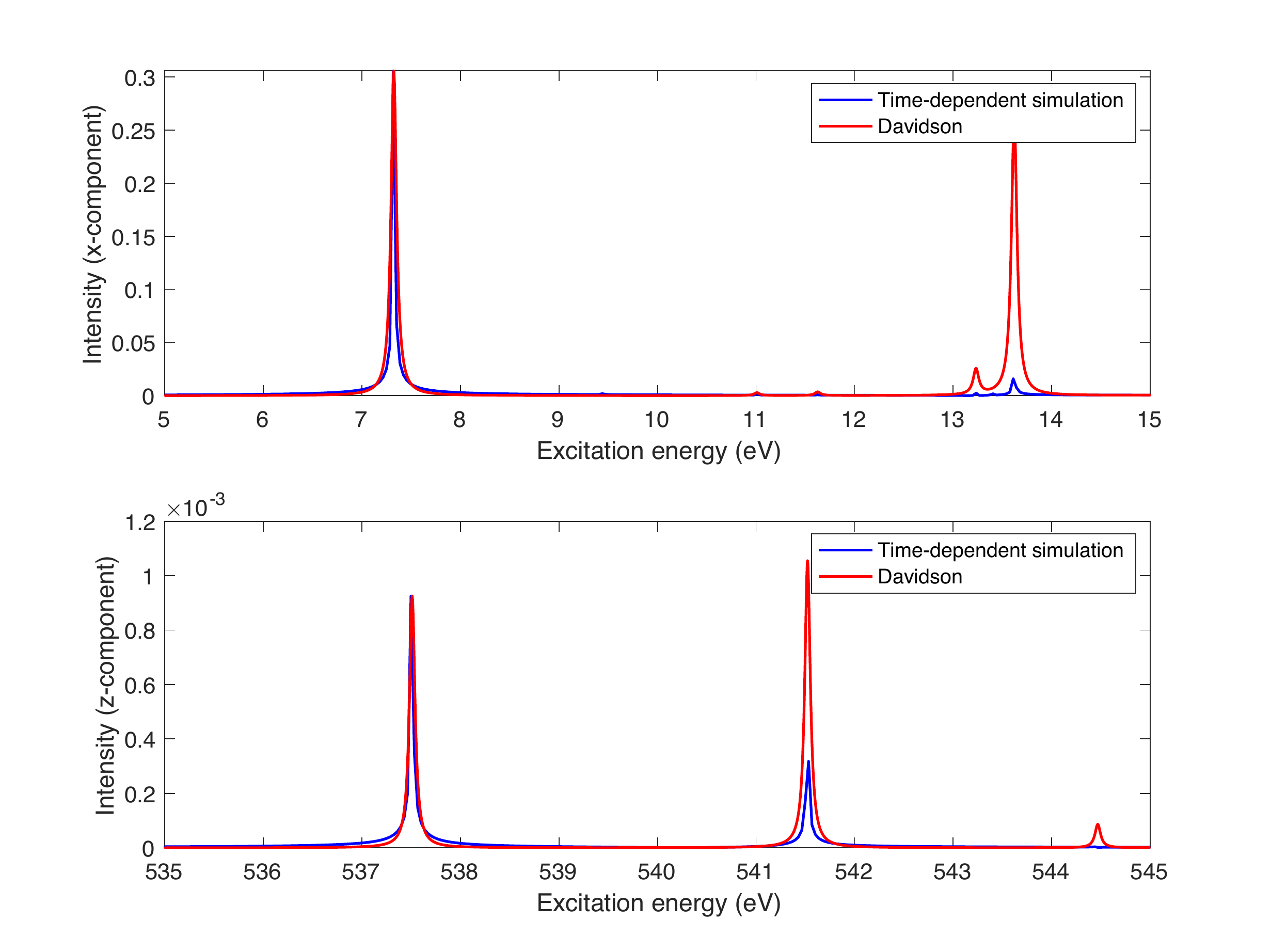}
    \caption{Water UV and X-ray CCSD absorption spectra obtained 
    using Davidson (CVS/aug-cc-pCVDZ for X-ray, aug-cc-pVDZ for UV) 
    and real-time propagation
    (aug-cc-pCVDZ for X-ray, aug-cc-pVDZ for UV). 
    The top and bottom plots show simulated UV and X-ray spectra, 
    respectively. 
    The Davidson spectra were produced by applying Lorentzian broadening to the stick spectra ($\SI{0.0025}{\hartree}$~FWHM). Intensities from the time-dependent simulation have been scaled 
    so that the intensity of the first peak matches the EOM oscillator strength. 
    }
    \label{fig:td_h2o}
\end{figure}

Spectroscopic properties can also be modelled with the Lanczos method or with real-time propagation of the coupled cluster wave function.
In Figure \ref{fig:valence_h2o}, 
we show CCSD/aug-cc-pCVDZ UV/Vis absorption spectra of H$_2$O,\citep{Olsen1996}
calculated using the Davidson (top) and  asymmetric Lanczos (bottom) algorithms. Note that we have artificially extended the spectra 
beyond the ionization potential ($\SI{12.3}{\eV}$ IP-CCSD/aug-cc-pCVDZ) to illustrate convergence behavior. With the Lanczos algorithm, 
the low energy part of the spectrum converges with a smaller reduced space 
than the high energy part.\citep{Coriani2012}

We have also generated oxygen edge X-ray absorption spectra using the 
Davidson 
and Lanczos algorithms with CVS projection; 
see Figure \ref{fig:cvs_h2o}. 
We see the same overall behavior as in Figure \ref{fig:valence_h2o}. 

Absorption spectra can also be obtained from real-time propagation of the coupled cluster wavefunction. 
See Figure \ref{fig:td_h2o} for UV/Vis and oxygen edge X-ray absorption spectra.
See the supplemental material for computational details.
The first peak in both plots has been scaled 
to match the intensity obtained using Davidson. 
The position of the peaks are the same with both approaches, but the intensities differ because we specified pulses with frequency distributions centered on the first excitation energy.

\section{Concluding remarks}
\eTone~
is an optimized open source electronic structure program.
Several features are worth emphasizing.
To the best of our knowledge, 
our CC3 implementation is the fastest for calculating ground and excited state energies and EOM oscillator strengths.
The low memory CC2 code has memory and disk requirements of order $\mathcal{O}(n_{\mrm{MO}}^2)$ 
and $\mathcal{O}(n_{\mrm{MO}}^3)$, respectively, allowing us to treat systems with thousands of basis functions.
At the core of our program is the Cholesky decomposition of the electron repulsion integral matrix;
our implementation is faster 
and less storage intensive than that of any other program.
Exciting new developments are also part of \eT. 
It features the only spin adapted closed shell implementation of time-dependent coupled cluster theory.
Furthermore, 
the MLHF and MLCC2 methods 
extend the treatable system size without sacrificing accuracy 
for intensive properties such as excitation energies.

The \eT~source code is written in modern object oriented Fortran, 
making it easy to expand and contribute to the program.
It is freely available on GitLab\citep{etGitlab} 
and the manual can be found at www.etprogram.org.
We will continue to expand the capabilities of \eT, 
focusing on molecular properties and multilevel methods.
We believe the program will be useful for the 
quantum chemistry community,
both as a development platform 
and for production calculations.

\section{Acknowledgements}
We thank Sander Roet for 
useful discussions regarding
Python and Git functionality.
We thank Franco Egidi, Laura Grazioli, Gioia Marrazzini, Rosario Roberto Riso, and Anna Kristina Schnack-Petersen, who attended the \eT~workshop in Pisa, 
October/November 2019. 
We also thank Edward Valeev for assistance with Libint and Roberto Di Remigio
for help with a patch of PCMSolver.
We acknowledge computing resources through UNINETT Sigma2 - 
the National Infrastructure for High Performance Computing and Data Storage in Norway,
through project number NN2962k and NN9409k, and the SMART@SNS Laboratory. 
We acknowledge funding from the Marie Sk{\l}odowska-Curie European Training Network ``COSINE - COmputational Spectroscopy In Natural sciences and Engineering'', Grant Agreement No. 765739, the Research Council of Norway through FRINATEK projects 263110 and 275506. S.C. acknowledges support from the Independent Research Fund Denmark -- DFF-RP2 grant no. 7014-00258B

\bibliography{eTbibs}

\end{document}